%% file: arxiv.tex
\documentclass[%
 reprint,
superscriptaddress,
footinbib,
amsmath,amssymb,
pre,
nobibnotes
]{revtex4-1}

\usepackage[english]{babel}
\usepackage[utf8x]{inputenc}
\usepackage[colorinlistoftodos]{todonotes}
\usepackage{enumerate}
\usepackage{graphicx}
\usepackage{dcolumn}
\usepackage{bm}

\usepackage{textcomp}
\usepackage[pagewise,columnwise]{lineno}
\usepackage{skak}

\usepackage[colorinlistoftodos]{todonotes}

\begin{document}

\preprint{APS/123-QED}

\title{Assessing the effectiveness of barrier allocation strategies against the propagation of phytopathogens and pests with percolation
}

\author{E. G. García Prieto}
\affiliation{Facultad de Ciencias F\'isico Matem\'aticas, Benem\'erita Universidad Aut\'onoma de Puebla, Apartado Postal 165, 72000 Puebla, Pue., M\'exico}

\author{G. García Morales}
\affiliation{Facultad de Ciencias F\'isico Matem\'aticas, Benem\'erita Universidad Aut\'onoma de Puebla, Apartado Postal 165, 72000 Puebla, Pue., M\'exico}

\author{J. D. Silva Montiel}
\affiliation{Facultad de Ciencias F\'isico Matem\'aticas, Benem\'erita Universidad Aut\'onoma de Puebla, Apartado Postal 165, 72000 Puebla, Pue., M\'exico}

\author{D. Rosales Herrera}
\affiliation{Facultad de Ciencias F\'isico Matem\'aticas, Benem\'erita Universidad Aut\'onoma de Puebla, Apartado Postal 165, 72000 Puebla, Pue., M\'exico}

\author{J. R. Alvarado García}
\affiliation{Facultad de Ciencias F\'isico Matem\'aticas, Benem\'erita Universidad Aut\'onoma de Puebla, Apartado Postal 165, 72000 Puebla, Pue., M\'exico}

\author{A. Fern\'andez T\'ellez}
\affiliation{Facultad de Ciencias F\'isico Matem\'aticas, Benem\'erita Universidad Aut\'onoma de Puebla, Apartado Postal 165, 72000 Puebla, Pue., M\'exico}

\author{Y. Martínez Laguna}
\affiliation{Vicerrectoría de Investigación y Estudios de Posgrado, Benemérita Universidad Autónoma de Puebla, Apartado Postal 165, 72000 Puebla, Pue., México}

\author{J. F. López-Olguín}
\affiliation{Herbario y Jardín Botánico, Vicerrectoría de Investigación y Estudios de Posgrado, Benemérita Universidad Autónoma de Puebla, Apartado Postal 165, 72000 Puebla, Pue., México}
\affiliation{Centro de Agroecología,
Instituto de Ciencias,
Benemérita Universidad Autónoma de Puebla, Apartado Postal 165, 72000 Puebla, Pue., M\'exico}

\author{J. E. Ram\'irez}
\email{jhony.ramirezcancino@viep.com.mx}
\affiliation{Centro de Agroecología,
Instituto de Ciencias,
Benemérita Universidad Autónoma de Puebla, Apartado Postal 165, 72000 Puebla, Pue., M\'exico}

\begin{abstract}
We investigate the connectivity properties of square lattices with nearest-neighbor interactions, where some sites have a reduced coordination number, meaning that certain sites can only connect through three or two adjacent sites. 
This model is similar to the random placement of physical barriers in plantations aimed at decreasing connectivity between susceptible individuals, which could help prevent the spread of phytopathogens and pests. 
In this way, we estimate the percolation threshold as a function of the fraction of sites with a reduced coordination number ($p_d$), finding that the critical curves can be well described by a $q$-exponential function. 
Additionally, we establish the correlation between $p_d$ and the fraction of barriers effectively placed, which follows a power law behavior.
The latter is helpful in estimating the relative costs of the barrier allocation strategies. In particular, we found that the allocations of two barriers per site model $\{ \mathbin{\rotatebox[origin=c]{270}{$\with$}}, \mathbin{\rotatebox[origin=c]{90}{$\with$}} \}$ can produce savings between 5\% and 10\% of the strategy cost compared to the independently random barrier allocations (joint site-bond percolation).
From an agroecology perspective, adding barriers to the plantation gives farmers the opportunity to sow more vulnerable plant varieties.
\end{abstract}
\maketitle


\section{Introduction}

One of the key challenges in agroecology is preventing the spread of phytopathogens and pests that lead to enormous economic losses by using chemical-free and ecofriendly methods.
Among all phytopathogens and pests, the genus \emph{Phytophthora} (which literally means "plant destroyer" in Greek) stands out because of the \emph{Phytophthora} zoospores' motility, which can chemotactically disperse toward the neighboring plants through water films or soil moisture \cite{erwin1996phytophthora, bernhardt1982effect}.
Damages produced by \emph{Phytophthora} include rotting in seedlings, tubers, corms, the base of the stem, and other organs, mainly affecting the roots of many plant species \cite{rinehart2007concepts, lamour2013phytophthora}.
Due to the unique physiology of \emph{Phytophthora} zoospores, there are no effective chemical treatments to prevent the proliferation and spread of this phytopathogen, rendering necessary the research and development of nonchemical strategies that minimize or eliminate the propagation of \emph{Phytophthora} or other phytopathogens and pests that also spread from one plant to neighboring ones \cite{parra2001resistance,naqvi2024,Anomalous}.

In particular, the spread of \emph{Phytophthora} can be modeled as a percolation problem when the plantation meets specific conditions, such as a regular lattice layout, controlled soil moisture, and selecting the planting spacing as the maximum distance that the zoospores can travel before starving or dying by inanition \cite{chaos, ramirez2020site}.
In this case, the spread happens only to the nearest neighbor plants, resembling the problems studied by percolation theory \cite{Efros,stauffer,  SABERI20151, D'Souza03072019, Sahimi}.

It was found in experimental trials that some individuals (of the same plant variety) can exhibit defense mechanisms against the infestation by a specific \emph{Phytophthora} strain, but there is no chemical or physical methodology to identify which seeds will grow as a resistant plant \cite{cohen1986systemic, erwin1996phytophthora, GILARDI201313, JE}.
Therefore, the plant-pathogen interaction provides another relevant parameter, the susceptibility ($\chi$), which can be interpreted as the probability of an individual getting ill after exposure to the phytopathogen.
Here, the spread of the illness only occurs through susceptible plants, which are uniformly distributed within the plantation.
In this way, in a monoculture plantation, the outbreak occurs if the susceptibility surpasses a threshold value that allows the formation of a spanning cluster (a large cluster connecting two opposite sides of the plantation), the well-known \emph{percolation threshold} \cite{stauffer, chaos, alonso2025applications}.
In the particular case of plantations disposed in square lattice arrangements where the phytopathogens or pests can only spread over the nearest neighbor plant and very low densities of infected cells at the beginning of the propagation process, the critical susceptibility is $\chi_c$=0.59274621(13)  \cite{newman2000efficient,Jacobsen, LI20211,Anomalous}.

This means that farmers should sow plant varieties with susceptibilities lower than $\chi_c$ to prevent the effects produced by the proliferation of phytopathogens or pests.
Nevertheless, there exist strategic crops or plant varieties with great commercial interest for which the susceptibility to a specific phytopathogen or pest is $\chi=1$.
In such cases, it is mandatory to reduce the connectivity of the network of susceptible plants to prevent an epidemic scenario \cite{callaway2000network, PhysRevE.70.066145}.
One agroecological strategy proposed to achieve connectivity reduction is to diminish plantation density by leaving empty cells randomly distributed across the plantation, similarly to the problems studied in percolation theory \cite{JE}.
Although this strategy may control the spread of the disease, the decreasing total yield production could be unprofitable from the farmer's point of view.

Another effective agroecological strategy is implementing polyculture systems \cite{MARTING,ELHA}, where the introduction of a second resistant plant variety can reduce the connectivity of the plantation \cite{chaos}. 
Here, the resistant individuals act as physical barriers that locally prevent the spread of phytopathogens and pests. Consequently, if the density and susceptibility of the second variety are sufficient, the epidemic can be prevented as a collective phenomenon originated from resistant plants.
However, this approach is limited to the initial design of the plantation. Hence, the intercropping is suitable for crops with a relatively short producing life \cite{JE, chaos}.

In contrast, long-lived plantations that produce for extended periods require a different approach to polyculture systems. For these cases, an alternative strategy to limit plantation connectivity is to install physical barriers, such as root barriers or trenches filled with manure, to hide neighboring plants, which may prevent spread in the direction where the barrier is placed \cite{ramirez2020site}.

In this paper, we aim to analyze the effectiveness of different barrier allocation configurations from the perspective of percolation theory.
In this way, we assess the percolation threshold (critical susceptibility) of random square lattices with nearest neighbor interaction wherein a fraction of sites has a reduced coordination number, denoted by $p_d$. 
Since the barriers are oriented perpendicular to the local spreading direction, the reduction in the coordination number resembles installing a barrier that disconnects two neighboring plants.
In Table~\ref{tab:allocations}, we illustrate four different arrangements of random barrier allocations by placing one or two barriers on sites with reduced coordination number.
Therefore, the most effective strategy will be the one requiring the lowest number of barriers to prevent the outbreak.
For simplicity in our computational implementation, we consider fully sown plantations and a very low percentage of inoculated cells at the beginning of the propagation process. 
The latter condition is used to prevent additional edges caused by the motility of the phytopathogens \cite{Herrera2024}.
Under these conditions, the percolation threshold must be understood as the maximum susceptibility at which the outbreak is avoided.

The rest of this paper is organized as follows.
In Sec.~\ref{sec:method}, we discuss our computational implementation of the simulations together with the data analysis method to estimate the critical susceptibility in the thermodynamic limit.
Section~\ref{sec:results} contains our results of the critical susceptibility as a function of the fraction of sites with reduced coordination number.
Additionally, we present the correlation between the fraction of sites with barriers and the effective fraction of barriers allocated in the plantation.
This result is relevant because it allows us to evaluate the best barrier allocation strategy for preventing outbreaks.
Finally, we wrap up this work with our concluding remarks in Sec.~\ref{sec:conclusions}.

\input{taballocations}

\section{Simulation method and data analysis}\label{sec:method}

For the simulations, we adopt the microcanonical scheme developed by Newman and Ziff \cite{ziff2, newman2000efficient}.
In brief, this algorithm consists of measuring an observable $O$ when exactly $n$ sites are occupied in the system ($O_n$).
Therefore, the average value of $O$ at a certain occupation probability $p$ is computed as the convolution of $O_n$ with the fluctuations of the number of occupied sites at constant $p$, that is, 
\begin{equation}
    O(p)=\sum_n O_n \mathbb{P}_n(p). 
\end{equation}
In the particular case of uniform random distribution of occupied sites along with the lattice, the fluctuations correspond to the binomial probability mass distribution, given by 
\begin{equation}
    \mathbb{P}_n(p)=\mathcal{B}(N, n, p)= \binom{N}{n}p^n(1-p)^{N-n},
\end{equation}
where $N=L^2$ is the total number of sites in a square lattice of size $L\times L$.
The simulation is prepared by labeling each site at $i$-column and $j$-row according to the linearization $M=iL+j$.
The set of labels is very helpful for the clustering process, for which we use the Union-Find algorithm.
Since some sites have a reduced number of possible connection directions, we also require two additional arrays of length $L(L-1)$ to store the occupation state of the horizontal and vertical bonds connecting adjacent sites. 
Thus, if the $k$-entry is one, the bond is open and the clustering between the adjacent sites is possible. 
Otherwise, the $k$-entry is zero, meaning that the adjacent sites are disconnected.

We start the simulation by generating the number $n_d$ of sites with a reduced coordination number from the binomial distribution $B(L^2, p_d)$.
Then, $n_d$ labels are randomly taken from the set of sites' labels.
For each of these sites, one or two bonds are randomly selected to be closed depending on the model (see Table~\ref{tab:allocations}), and the corresponding entry is changed from one to zero in the horizontal or vertical bond arrays accordingly, indicating that these bonds are closed and they cannot connect the corresponding adjacent sites.

Next, we randomly and independently add sites one by one until a spanning cluster appears. 
In each step, we perform the clustering by checking the occupancy state of the neighboring sites and all the bonds connecting them.
The added site is merged to an adjacent cluster if, at the same time, the adjacent site is occupied and the corresponding bond is open.
The adding site process stops when the spanning cluster appears. Here, we record the critical number of sites needed to form the spanning cluster.

\begin{figure}
    \centering
    \includegraphics{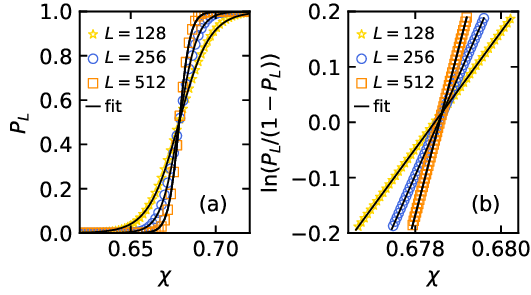}
    \caption{(a) Characteristic sigmoidal behavior of the percolation probability. (b) Linear behavior of the function $\ln(P_L/(1-P_L))$ around the percolation threshold. For this example, we use the data simulated for the case sq2N-2 with $p_d=$0.175.}
    \label{fig:PP}
\end{figure}

\begin{figure}
    \centering
    \includegraphics{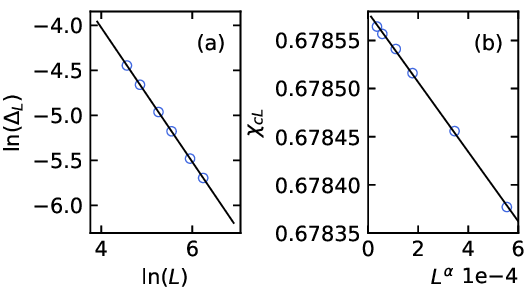}
    \caption{Finite size effects of (a) $\Delta_L$ and (b) $\chi_{cL}$. Open circles are the data obtained by analyzing the data from simulations. Solid lines correspond to the scaling laws $\Delta_L\propto L^{-1/\nu}$ and $p_c-p_{cL}\propto L^\alpha$ for the width transition and the percolation threshold, respectively. For this example, we use the data simulated for the case sq2N-2 with $p_d=$0.175.}
    \label{fig:sc}
\end{figure}

With the information from the simulations, we estimate the probability $f(n)$ of observing the emergence of the spanning cluster after adding exactly $n$ sites as the ratio of the frequency of simulations stopped after adding $n$ sites to the total number of simulations.
Thus, the cumulative sum 
\begin{equation}
    F(n)=\sum_{k=0}^nf(k)
\end{equation}
is the probability of observing the emergence of the spanning cluster after adding at most $n$ sites.
Therefore, the percolation probability at a susceptibility $\chi$ is estimated as 
\begin{equation}
   P_L(\chi)=\sum_{n=0}^{L^2}F(n) \mathcal{B}(L^2,n, \chi ). 
   \label{eq:percprob}
\end{equation}
The subscript $L$ denotes an implicit dependence of the percolation probability on the system size.
In general, the estimation of the $P_L$ requires the computation of the factorial of large numbers, which may represent a numerically difficult task.
We turn this problem around by recursively computing the binomial weights by using the following formula \cite{ziff2}:
\begin{eqnarray}
    B(L^2,n,\chi)= \left\{
    \begin{array}{lcc}
   		B(L^2,n-1,\chi)\frac{L^2-n+1}{n}\frac{\chi}{1-\chi}& \mathrm{if} & n>n_{\text{m}},\\
		\\
		B(L^2,n+1,\chi)\frac{n+1}{L^2-n}\frac{1-\chi}{\chi}& \mathrm{if} & n<n_{\text{m}},
			\end{array}
\right. \nonumber
\label{eq:recursiv-binomial}
\end{eqnarray}
where $n_{\text{m}}=\chi L^2$ is the $n$-value where the probability mass function of the binomial distribution takes its maximum value. 
Additionally, we set $B(L^2,n_\text{m},\chi)=1$.
In this way, the percolation probability \eqref{eq:percprob} must be normalized by dividing by $\sum_{n=1}^{L^2}B(L^2,n,\chi)$.

It is well-known that the percolation probability behaves as a sigmoidal curve (see Fig.~\ref{fig:PP} (a)), which can be fitted with the function 
\begin{equation}
    g(\chi)=\frac{1}{2}\left[1+\tanh\left(\frac{\chi-\chi_{cL}}{\Delta_L}\right) \right],
\end{equation}
where $\chi_{cL}$ and $\Delta_L$ are the estimation of the percolation threshold for finite systems and the width of the percolation transition, respectively \cite{Rintoul}.
In this way, the percolation threshold for finite systems satisfies the condition $P_L(\chi_{cL})=1/2$.

We made a first rough estimation of $\chi_{cL}^*$ and $\Delta_L^*$ by computing the percolation probability in the susceptibility range [0.1,1.0] and fitting the $g(\chi)$ function to the generated data.
This estimation can be improved by restricting the range wherein $P_L(\chi)$ is computed.
We select the range [$\chi_-$, $\chi_+$], where 
\begin{equation}
   \chi_{\pm}=\chi_{cL}^*+\frac{\Delta_L^*}{2}\ln\left( \frac{1\pm2\epsilon}{1\mp 2\epsilon} \right). 
\end{equation}
In this range, the percolation threshold takes values around 0.5 for low values of $\epsilon$.
Additionally, under these conditions, the function $\ln(P_L(\chi)/(1-P_L(\chi)))$ behaves as a linear function (as depicted in Fig.~\ref{fig:PP} (b)), where the slope and the intercept are $1/\Delta_L$ and $-\chi_{cL}/\Delta_L$, respectively.
By applying this methodology, we obtain a better determination of $\chi_{cL}$.

Usually, the observables in percolation theory exhibit a dependency on the system size as we depict in Fig.~\ref{fig:sc}, which is described by scaling laws \cite{stauffer, Sahimi,Aconiglio_1982}. In particular, the width of the percolation transition and the percolation threshold satisfy
\begin{align}
    \Delta_L \propto & L^{-1/\nu}, \label{eq:scDL}\\
    \chi_c-\chi_{cL} \propto & L^\alpha, \label{eq:scpc}
\end{align}
respectively \cite{Dolz2005}.
In Eq.~\eqref{eq:scDL}, $\nu$ corresponds to the exponent associated with the correlation length, taking values around 4/3 for all simulations, which is in agreement with the results reported in the literature for percolation theory in two-dimensions.
In Fig.~\ref{fig:sc} (a), we show an example of this scaling law in our simulations.
On the logarithmic scale, the estimation of $\nu$ is done by determining the slope of the trend of $\Delta_L$ as a function of system size in a log-log plot.

Additionally, Eq.~\eqref{eq:scpc} provides a methodology to estimate the percolation threshold on the thermodynamic limit by extrapolating the trend of $\chi_{cL}$ as a function of $L^{\alpha}$.
Here, $\alpha$ is a negative exponent indicating the convergence of $\chi_{cL}$ as $L$ increases.
Moreover, with the adequate value of $\alpha$, $\chi_{cL}$ is a linear function of $L^\alpha$, as we depict in Fig.~\ref{fig:sc} (b).
In all the cases analyzed in this paper, we found $-2<\alpha<-1.5=-2/\nu$.
These results indicate that our simulations offer a more restrictive scaling law for the percolation threshold compared to other methods, such as the Hoshen-Kopelman algorithm, for which $\alpha=-1/\nu$ \cite{hoshen1976, Dolz2005}.
Thus, we can improve the accuracy of the estimation of the percolation threshold in the thermodynamic limit.
Since $\alpha$ takes negative values, the percolation threshold in the thermodynamic limit ($L\to\infty$) is estimated as the intercept of the linear trend of Eq.~\eqref{eq:scpc}.
Figure~\ref{fig:sc} (b) shows an example of the determination of the percolation threshold through the extrapolation of the finite size effects described above. 

\section{Results}\label{sec:results}

\begin{figure*}
    \centering
    \includegraphics{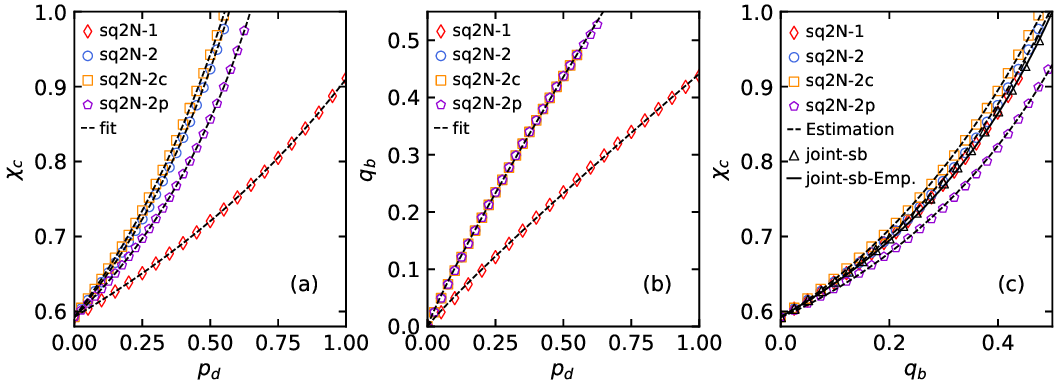}
    \caption{(a) Critical susceptibility (figures) as a function of the fraction of sites with a reduced coordination number ($p_d$). Dashed lines correspond to the $q$-exponential fitting in Eq.~\eqref{eq:chipd}. (b) Correlation between the fraction of sites with a reduced coordination number and the fraction of the barriers effectively paced. Note that the curve $q_b=q_b(p_d)$ collapses into a single trend for all cases wherein there are two removed edges. Dashed lines correspond to the power law fitting in Eq.~\eqref{eq:qbpd}. (c) Critical susceptibility (figures) as a function of the fraction of the barriers effectively placed ($q_b$). Additionally, we also plot the results of the percolation threshold for the joint site-bond percolation model (red triangles). Dashed lines correspond to the estimation of the critical susceptibility as a function of $q_b$. Solid line corresponds to the empirical critical curve for the joint site-bond percolation model reported in Ref.~\cite{Tarasevich1}.}
    \label{fig:res}
\end{figure*}

\begin{table}
\caption{Fitted values of the parametrizations in Eqs.~\eqref{eq:chipd} and \eqref{eq:qbpd}.}
\centering

\begin{tabular}{c c c c c}
\hline
Model & $\lambda$ & $q$ & $\sigma$ & $\tau$ \\ \hline
sq2N-1 &  0.360(3)  &  0.153(6) & 0.441(1)  & 0.923(5)  \\
sq2N-2 & 0.7262(3)   &  0.0687(3) &  0.816(4) & 0.903(6)  \\
sq2N-2c &  0.801(7)  & 0.351(4)  &  0.816(4) & 0.903(6)  \\
sq2N-2p &  0.585(2)  &  -0.304(3) &   0.816(4) & 0.903(6)  \\ \hline
    \end{tabular}%
\label{tab:fit_values}
\end{table}

In this section, we present our results of the critical susceptibility obtained by using the simulation method and data analysis discussed in Sec.~\ref{sec:method}. 
For each estimation in the thermodynamic limit, we performed 10$^7$ simulations for the system size $L=$ 96, 128, 192, 256, 384, and 512.
For all models of barrier allocations presented in Table~\ref{tab:allocations}, we begin by exploring the case without sites with reduced coordination number ($p_d=0$), which resembles the purely site percolation of square lattices with nearest neighbor interactions. 
In this case, we obtain $\chi_c=0.592746(2)$, which is in agreement with the best estimation of the site percolation threshold for the square lattice \cite{chaos,stauffer}. Later, the value of $p_d$ is increased by $\Delta p_d=$0.05 and 0.025 until reaching the susceptibility value $\chi_c=$1 for the models with one and two edges removed, respectively.

Figure~\ref{fig:res} (a) shows our results of the critical susceptibility as a function of the fraction of sites with reduced coordination number ($p_d$) for all the models of barrier allocations (see Table~\ref{tab:allocations}).
In all cases, the critical susceptibility is a monotonically increasing function of $p_d$, indicating that the allocation of barriers decreases the plantation connectivity and therefore allows the possibility of sowing more susceptible plant varieties.
Interestingly, the critical curves $\chi_c=\chi_c(p_d)$ can be well fitted by the following function:
\begin{equation}
    \chi_c=\frac{p_{cs}}{e_q(-\lambda p_d)},\label{eq:chipd}
\end{equation}
where $p_{cs}=$0.59274621(13) is the percolation threshold for square lattices with nearest neighbor interactions and $e_q(x)$ denotes the $q$-exponential function \cite{abe, YAMANO2002486}, defined as
\begin{equation}
    e_q(x)=\left(1+(1-q)x\right)^{\frac{1}{1-q}},
\end{equation}
which is widely used for describing nonextensive phenomena in statistical physics, representing a generalization of the Boltzmann-Gibbs statistics \cite{tsallis1988}.
Table~\ref{tab:fit_values} contains the fitted values of $q$ and $\lambda$ for all the barrier allocation models discussed in this manuscript.
In particular, note that $q<1$ in all cases, meaning that the $q$-exponential function in \eqref{eq:chipd} has a bounded domain \cite{wilk2000,budini,budini2}, explicitly, the support is $[0, 1/\lambda(1-q))$ but $\lambda(1-q)<1$ in all cases, and thus, Equation~\eqref{eq:chipd} is well defined for $0\leq p_d \leq 1$.

Additionally, the series expansion of Eq.~\eqref{eq:chipd} around $p_d=0$ leads to
\begin{equation}
    \chi_c \approx p_{cs}(1+\lambda p_d) \approx p_{cs} e^{\lambda p_d}.
\end{equation}
The latter result means that the inclusion of a low number of sites with barriers exponentially modifies the critical susceptibility, allowing the producer to protect their plantation against the outbreaks. This effect can be magnified by introducing more barriers,  where the critical susceptibility grows faster than an exponential function.

We want to emphasize that our model considers the random allocation of sites with a reduced coordination number, which can be understood as the inclusion of barriers in the plantation. 
However, two neighboring sites of this kind can share the same removed edge, resulting in a decrease in the effective number of barriers placed.
In what follows, we denote the fraction of effective barriers placed as $q_b$.
To establish the correlation between $p_d$ and $q_b$, in the simulations we count the number of new barriers placed, ignoring those already allocated.
We found that the number of barriers effectively placed for fixed $p_d$ follows a normal distribution less wide than a binomial distribution.
Therefore, we compute $q_b=\langle N_b \rangle/2L(L-1)$, where $\langle N_b \rangle$ is the average number of barriers effectively placed and $2L(L-1)$ is the total edges in the lattices.
In Fig.~\ref{fig:res} (b), we show the results obtained for $q_b$ as a function of $p_d$.
Note that $q_b$ and $p_d$ are not linearly correlated, and are better described by a power law
\begin{equation}
    q_b=\sigma p_d^\tau.
    \label{eq:qbpd}
\end{equation}
In Table~\ref{tab:fit_values}, we report the fitted values of $\sigma$ and $\tau$ for all the models of barrier allocations discussed in this manuscript.
As expected, the fraction of barriers effectively placed collapses into a single curve for the models labeled as sq2N-2, sq2N-2 parallels, and sq2N-2 corners, since in all these cases, two edges were removed per site with reduced coordination number. 

Equation~\eqref{eq:qbpd} is relevant because it allows us to compare the performance of the different barrier allocation models as a function of the fraction of the barriers effectively placed. 
This is done by writing down $p_d$ as a function of $q_b$ from Eq.~\eqref{eq:qbpd}, and substituting the resulting expression in Eq.~\eqref{eq:chipd}.
Figure~\ref{fig:res} (c) shows our results of the critical susceptibility as a function of $q_b$.

\begin{figure}
    \centering
    \includegraphics{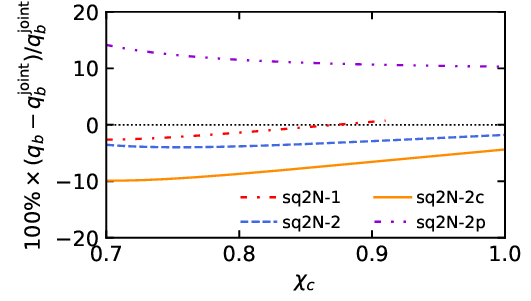}
    \caption{Relative costs of the barrier allocations with respect to the independently random barrier allocations (joint site-bond percolation model). Negative values of the relative cost mean savings for the farmers. The dotted line corresponds to the zero value of the relative cost, plotted as a guide.}
    \label{fig:comparacion}
\end{figure}

Another comparison we can make is with the joint site-bond percolation model, which has been proposed as a model for barrier allocations to prevent the propagation of \emph{Phytophthora} on plantations with long producing life \cite{ramirez2020site}.
In this model, the edges are independently open with probability $p_b$, or equivalently, the fraction of barriers placed is $q_b=1-p_b$.
Here, one problem to solve is the determination of the critical value of $p_b$ (or $q_b$) required for the formation of a spanning cluster of sites with occupation probability $p_s$.
This model can be understood as a generalization of the site and bond percolation, which are recovered in the cases of considering $p_b=1$ and $p_s=1$, yielding the percolation threshold for the purely site ($p_{cs}=0.59274621(13)$) and bond ($p_{cb}=0.5$) percolation, respectively.
In Fig.~\ref{fig:res} (c), we also include our results for the determination via simulations of the critical curve for the joint site-bond percolation.
Moreover, the critical curves of the joint site-bond percolation are empirically described in terms of the site ($p_s$) and bond ($p_b$) occupation probabilities by \cite{Tarasevich1, LEBRECHT2025130400}
\begin{equation}
    p_b=\frac{B}{p_s+A},
    \label{eq:joint}
\end{equation}
with
\begin{align}
    A=&\frac{p_{cb}-p_{cs}}{1-p_{cb}}, \nonumber \\ 
    B=&\frac{p_{cb}(1-p_{cs})}{1-p_{cb}}. \nonumber
\end{align}

Using the empirical critical curve \eqref{eq:joint}, we define the relative cost of the barrier allocation model by considering the joint site-bond percolation model as a baseline strategy, which is given by
\begin{equation}
    \eta(\chi_c)=100\%\frac{q_b(\chi_c)-q_b^\text{joint}(\chi_c)}{q_b^\text{joint}(\chi_c)},
\end{equation}
where $q_b(\chi_c)$ and $q_b^\text{joint}(\chi_c)$ are the fraction of barriers placed to reach the critical susceptibility $\chi_c$ for a particular model and the joint site-bond percolation, respectively.
Figure~\ref{fig:comparacion} shows our results of the relative costs of the barrier allocation models.
We must emphasize that negative values of the relative cost indicate the model requires fewer barriers to prevent an outbreak for a plant with a specific susceptibility level. 
According to our simulations, the models sq2N-2 and sq2N-2 corners require a smaller number of barriers to prevent an outbreak. It is worth mentioning that the model sq2N-2 corners has a range of savings between 5\% and 10\% compared to independently randomly allocated barriers.

\begin{figure*}
    \centering
    \includegraphics{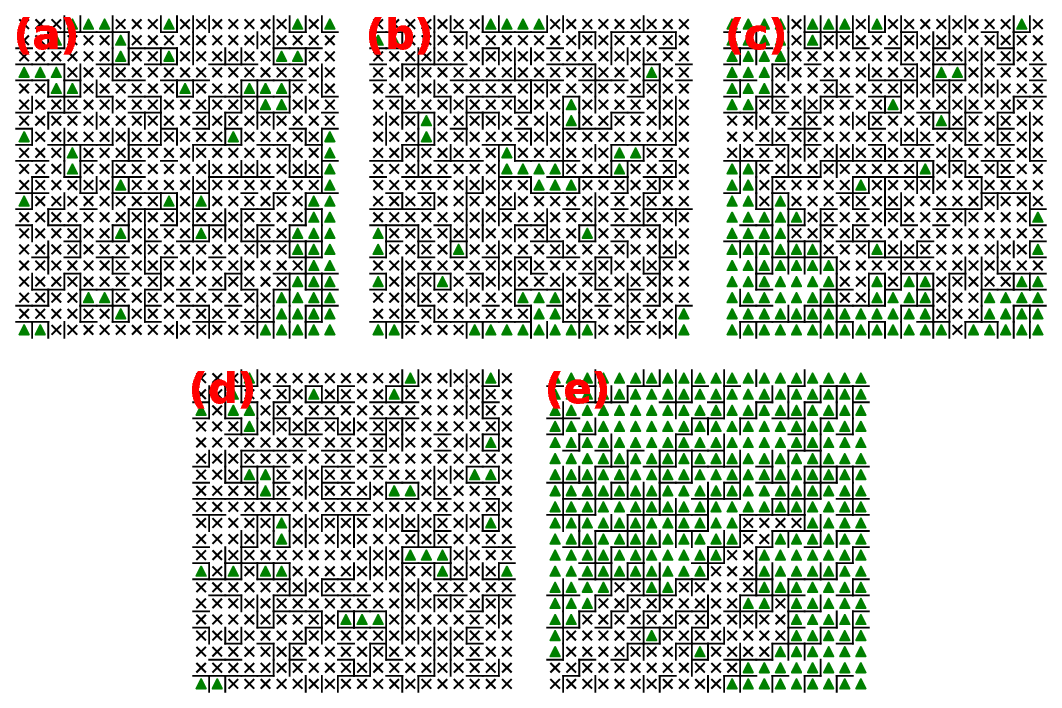}
    \caption{Largest cluster (crosses) of susceptible plants under the barrier allocation models: (a) joint site-bond percolation, (b) sq2N-1, (c) sq2N-2, (d) sq2N-2 parallels, and (e) sq2N-2 corners. In the simulations, we consider $\chi=1$ and $q_b=0.475$. Note the capability of the model sq2N-2 corners to fragment the largest cluster for lower densities of the fraction of barriers effectively placed than the joint site-bond percolation model. Therefore, in this case, the propagation of phytopathogens and pests will only occur on finite clusters.}
    \label{fig:samples}
\end{figure*}

\section{Concluding remarks}\label{sec:conclusions}

In this work, we introduced the model of percolation systems where a fraction of sites have a reduced coordination number. 
This means that in a square lattice with nearest neighbor interactions, some sites can be connected only through 3 or 2 edges, depending on the models studied, instead of the 4 edges in the traditional percolation model.
From an agroecological perspective, these sites with a reduced coordination number can be understood as a reduction in plantation connectivity achieved by adding physical barriers, such as root barriers or trenches filled with manure.
This model is inspired by the design of agroecological strategies that prevent the spread of \emph{Phytophthora} zoospores on long-lived plantations where the plantation layout cannot be changed between production cycles \cite{ramirez2020site}.

In Table 1, we illustrated the models of percolation that consider sites with a reduced coordination number with probability $p_d$. 
In the model sq2N-1, we removed one edge per site; meanwhile, two edges are removed in the models sq2N-2, sq2N-2 parallels, and sq2N-2 corners.
For each model, we estimated the percolation threshold in the thermodynamic limit as a function of the fraction of sites with a reduced coordination number.

In all cases, adding physical barriers to plantations allows for planting more susceptible plant varieties. 
In fact, for low values of $p_d$, the critical susceptibility increases exponentially from the percolation threshold of the purely site problem. 
This effect becomes even more significant for larger values of $p_d$, where the critical susceptibility increases faster than an exponential function. 
The main implication of our results is that adding barriers may help protect the plantation from the spread of phytopathogens and pests.
Additionally, the critical curves can be well described by a $q$-exponential function, which is a characteristic distribution in nonextensive statistical mechanics that reveals the complexity of this problem.

We also established a correlation between the fraction of sites with a reduced coordination number and the density of barriers effectively placed. Since two neighboring sites can share the same removed edge, the relationship between $p_d$ and $q_b$ is not linear, but rather follows a power-law behavior. 
Furthermore, this relationship helps us to describe the critical susceptibility as a function of the fraction of barriers effectively placed.

Interestingly, describing the critical susceptibility as a function of the fraction of barriers effectively placed allows for comparing the relative costs of barrier allocation strategies by examining the number of barriers needed to reach a specific susceptibility level.
To this end, we considered the joint site-bond percolation model as a baseline for estimating the relative cost of the strategy.
We discovered that the optimal strategy is the sq2N-2 corners model, which yields savings of 5\% to 10\% compared to the random barrier allocation.
In Fig.~\ref{fig:samples}, we show samples of the largest cluster formed in plantations (crosses in Fig.~\ref{fig:samples}) with the barrier strategies discussed in this manuscript (see Table~\ref{tab:allocations}) at $q_b=0.475$.
Note that the ability of the sq2N-2 corners model to fragment the largest cluster, assuring that in case of an epidemic episode, the dissemination will occur only on a finite cluster (a small part of the plantation).

We would like to conclude this paper by highlighting that barrier allocation strategies can be beneficial in protecting plantations against the propagation of phytopathogens and pests. However, it has been demonstrated that the intercropping system can enhance the overall cost-effectiveness of the strategy, leading to greater savings for farmers.

\begin{acknowledgments}
This work was funded by Secretaria de Ciencia, Humanidades, Tecnología e Innovación (SECIHTI-México) under the project CF-2019/2042, graduated and postdoctoral fellowships (grant numbers 1140160, 645654, and 289198).
We also acknowledge financial support from VIEP-BUAP under the project 2024/Exp123. 
\end{acknowledgments}

\bibliography{ref}

\end{document}

%% file: taballocations.tex
\begin{table}
\centering
\caption{Possible allocations of the barriers for the models of sites with a reduced coordination number. Dashed lines are the possible directions of propagation of \emph{Phytophthora} zoospores, which resemble the square lattices with nearest neighbor interactions. Solid lines correspond to the possible barrier allocations for the models considering a fraction of sites with a reduced coordination number.}
\label{tab:allocations}
    \begin{tabular}{ |c| c|}
    
    \hline
    Model &  Barrier allocations \\
    \hline
    \centering sq2N-1 &  
     \hspace{0.25cm}
\begin{tikzpicture}
    \draw[dashed](-0.5,0)--(0.5,0);
    \draw[dashed] (0,-0.5)--(0,0.5);
    \fill (0,0) circle(1pt);
    \draw[very thick, blue] (-0.25,-0.25)--(0.25,-0.25);
\end{tikzpicture}
 \hspace{0.25cm}
\begin{tikzpicture}
    \draw[dashed](-0.5,0)--(0.5,0);
    \draw[dashed] (0,-0.5)--(0,0.5);
    \fill (0,0) circle(1pt);
    \draw[very thick, blue] (-0.25,0.25)--(0.25,0.25);
\end{tikzpicture}
 \hspace{0.25cm}
\begin{tikzpicture}
    \draw[dashed](-0.5,0)--(0.5,0);
    \draw[dashed] (0,-0.5)--(0,0.5);
    \fill (0,0) circle(1pt);
    \draw[very thick, blue] (0.25,-0.25)--(0.25,0.25);
\end{tikzpicture}
 \hspace{0.25cm}
\begin{tikzpicture}
    \draw[dashed](-0.5,0)--(0.5,0);
    \draw[dashed] (0,-0.5)--(0,0.5);
    \fill (0,0) circle(1pt);
    \draw[very thick, blue] (-0.25,-0.25)--(-0.25,0.25);
\end{tikzpicture}  \hspace{0.25cm}\\

\hline
\centering sq2N-2  &
 \begin{tikzpicture}
    \draw[dashed](-0.5,0)--(0.5,0);
    \draw[dashed] (0,-0.5)--(0,0.5);
    \fill (0,0) circle(1pt);
    \draw[very thick, blue] (-0.25,-0.25)--(-0.25,0.25);
    \draw[very thick,blue] (-0.25,0.25)--(0.25,0.25);
\end{tikzpicture}
 \hspace{0.25cm}
\begin{tikzpicture}
    \draw[dashed](-0.5,0)--(0.5,0);
    \draw[dashed] (0,-0.5)--(0,0.5);
    \fill (0,0) circle(1pt);
    \draw[very thick, blue] (-0.25,-0.25)--(-0.25,0.25);
    \draw[very thick,blue] (-0.25,-0.25)--(0.25,-0.25);
\end{tikzpicture}
 \hspace{0.25cm}
\begin{tikzpicture}
    \draw[dashed](-0.5,0)--(0.5,0);
    \draw[dashed] (0,-0.5)--(0,0.5);
    \fill (0,0) circle(1pt);
    \draw[very thick, blue] (0.25,-0.25)--(0.25,0.25);
    \draw[very thick,blue] (-0.25,0.25)--(0.25,0.25);
\end{tikzpicture}
 \hspace{0.25cm}
\begin{tikzpicture}
    \draw[dashed](-0.5,0)--(0.5,0);
    \draw[dashed] (0,-0.5)--(0,0.5);
    \fill (0,0) circle(1pt);
    \draw[very thick, blue] (0.25,-0.25)--(0.25,0.25);
    \draw[very thick,blue] (-0.25,-0.25)--(0.25,-0.25);
\end{tikzpicture}  \hspace{0.25cm}\\
  &
\begin{tikzpicture}
    \draw[dashed](-0.5,0)--(0.5,0);
    \draw[dashed] (0,-0.5)--(0,0.5);
    \fill (0,0) circle(1pt);
    \draw[very thick, blue] (-0.25,-0.25)--(0.25,-0.25);
    \draw[very thick,blue] (-0.25,0.25)--(0.25,0.25);
\end{tikzpicture}
 \hspace{0.5cm}
\begin{tikzpicture}
    \draw[dashed](-0.5,0)--(0.5,0);
    \draw[dashed] (0,-0.5)--(0,0.5);
    \fill (0,0) circle(1pt);
    \draw[very thick, blue] (0.25,-0.25)--(0.25,0.25);
    \draw[very thick, blue] (-0.25,-0.25)--(-0.25,0.25);
\end{tikzpicture} \hspace{0.5cm}\\
\hline
 \centering sq2N-2 corners &
 \begin{tikzpicture}
    \draw[dashed](-0.5,0)--(0.5,0);
    \draw[dashed] (0,-0.5)--(0,0.5);
    \fill (0,0) circle(1pt);
    \draw[very thick, blue] (-0.25,-0.25)--(-0.25,0.25);
    \draw[very thick,blue] (-0.25,0.25)--(0.25,0.25);
\end{tikzpicture}
 \hspace{0.25cm}

 \hspace{0.25cm}
\begin{tikzpicture}
    \draw[dashed](-0.5,0)--(0.5,0);
    \draw[dashed] (0,-0.5)--(0,0.5);
    \fill (0,0) circle(1pt);
    \draw[very thick, blue] (0.25,-0.25)--(0.25,0.25);
    \draw[very thick,blue] (-0.25,-0.25)--(0.25,-0.25);
\end{tikzpicture} \hspace{0.25cm}\\
\hline
 \centering sq2N-2 parallels &
 \begin{tikzpicture}
    \draw[dashed](-0.5,0)--(0.5,0);
    \draw[dashed] (0,-0.5)--(0,0.5);
    \fill (0,0) circle(1pt);
    \draw[very thick, blue] (-0.25,-0.25)--(0.25,-0.25);
    \draw[very thick,blue] (-0.25,0.25)--(0.25,0.25);
\end{tikzpicture}
 \hspace{0.25cm}
\begin{tikzpicture}
    \draw[dashed](-0.5,0)--(0.5,0);
    \draw[dashed] (0,-0.5)--(0,0.5);
    \fill (0,0) circle(1pt);
    \draw[very thick, blue] (0.25,-0.25)--(0.25,0.25);
    \draw[very thick, blue] (-0.25,-0.25)--(-0.25,0.25);
\end{tikzpicture} \hspace{0.25cm}\\
    
\hline

    \end{tabular}
    
\end{table}